\documentclass{elsart}

 \usepackage{graphicx}

\usepackage{amssymb}

\begin{document}

\begin{frontmatter}



\title{Phase structure of a spherical surface model on fixed connectivity meshes}


\author[label1]{Hiroshi Koibuchi}
\ead{koibuchi@mech.ibaraki-ct.ac.jp}

\address[label1]{Department of Mechanical and Systems Engineering, Ibaraki National College of Technology, 
Nakane 866, Hitachinaka,  Ibaraki 312-8508, Japan}

\begin{abstract}
An elastic surface model is investigated by using the canonical Monte Carlo simulation technique on triangulated spherical meshes. The model undergoes a first-order collapsing transition and a continuous surface fluctuation transition. The shape of surfaces is maintained by a one-dimensional bending energy, which is defined on the mesh, and no two-dimensional bending energy is included in the Hamiltonian.    

\end{abstract}

\end{frontmatter}


\section{Introduction}\label{intro}
A considerable number of experimental studies have been conducted on the shape of membranes and their transformations \cite{Yoshikawa,Hotani}. One of the interesting topics in biological physics is to understand the origins of such a variety of shapes statistical mechanically on the basis of membrane models \cite{NELSON-SMMS2004-1,David-TDQGRS-1989}. The curvature model of Helfrich, Polyakov, and Kleinert \cite{HELFRICH-1973,POLYAKOV-NPB1986,KLEINERT-PLB1986} is well known as a model for describing shapes of membranes and has long been used for understanding the fluctuation of membranes \cite{Wiese-PTCP2000,Bowick-PREP2001,Gompper-Schick-PTC-1994,WHEATER-JP1994,KANTOR-NELSON-PRA1987,NELSON-SMMS2004-149}. From the two-dimensional differential geometrical view point, the curvature Hamiltonian is convenient to describe the mechanics of membranes \cite{NELSON-SMMS2004-149}.

However, the two-dimensional curvature Hamiltonian is not always necessary for providing the mechanical strength for the surface. In fact, it is well known that the cytoskeletal structures or the microtubules maintain shape of biological membranes \cite{Kusumi-BioJ-2004}.

One-dimensional bending energy can serve as the Hamiltonian for a model of membranes. Skeleton models are defined by using the one-dimensional bending energy, which is defined on a sub-lattice of a triangulated lattice \cite{KOIB-EPJB-2007,KOIB-JSTP2007,KOIB-PRE2007-2}. The compartmentalized structure constructed on the triangulated lattice is the sub-lattice and considered to be an origin of a variety of phases \cite{KOIB-PRE2007-2}. 

 The size of the sublattice is characterized by the total number $n$ of vertices inside a compartment. As a consequence, the mechanical strength of the surface varies depending on $n$, because the compartment size is proportional to $n$ and because the mechanical strength is given only by the sublattice. 

Therefore, it is interesting to see the dependence of the phase structure on $n$. The phase structure of the skeleton model is dependent not only on the bending rigidity $b$ but also on the size $n$. The phase structure of compartmentalized models at finite $n$ was partly studied as mentioned above \cite{KOIB-EPJB-2007,KOIB-JSTP2007,KOIB-PRE2007-2}. On the other hand, the models are expected to be in the collapsed phase in the limit of $n\!\to\!\infty$, because there is no source of the mechanical strength for the surface at $n\!\to\!\infty$.

However, the phase structure in the limit of $n\!\to\!0$ is unknown and yet to be studied. The compartmentalized model in this limit is governed by one-dimensional bending energy and, for this reason the model is expected be different from the model with the standard two-dimensional bending energy defined on triangulated surfaces without the compartments. 
 
In this Letter, we study a triangulated surface model defined by Hamiltonian that is a linear combination of the Gaussian bond potential and the one-dimensional bending energy. All the vertices are considered as junctions, which are considered to have a role for binding three one-dimensional chains. Consequently, the model is considered to be a compartmentalized model in the limit of $n\!\to\!0$. Note also that the model in this Letter is allowed to self-intersect \cite{GREST-JPIF1991,BOWICK-TRAVESSET-EPJE2001,BCTT-PRL2001}, and therefore the crumpled phase is expected to appear in the limit of $b\!\to\!0$, whereas the smooth phase is clearly expected in the limit of $b\!\to\!\infty$.

\section{Model and Monte Carlo technique}
\begin{figure}[hbt]
\centering
\includegraphics[width=6.0cm]{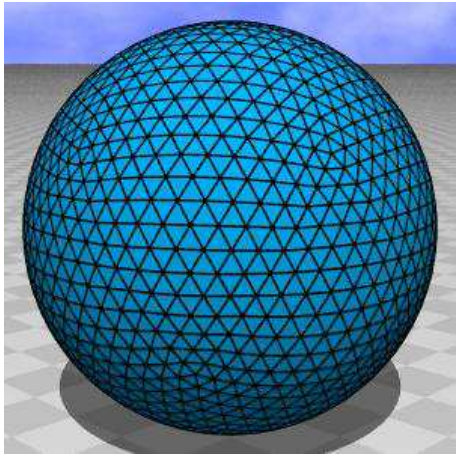}
\caption{A triangulated mesh of size $N\!=\!1442$, which is given by $\ell\!=\!12$. }
\label{fig-1}
\end{figure}

Triangulated meshes are obtained from the icosahedron such that the bonds are divided into $\ell$ pieces of the same length. Then we have the meshes of size $N\!=\!10\ell^2\!+\!2$, which include $12$ vertices of coordination number $q\!=\!5$ and, the remaining vertices are of coordination number $q\!=\!6$. Figure \ref{fig-1} shows the mesh of size $N\!=\!1442$, which is given by $\ell\!=\!12$. The triangle surfaces are shown in the figure in order to visualize the mesh more clearly than that without the triangle surfaces.

The Hamiltonian $S$ is given by a linear combination of the Gaussian bond potential $S_1$ and the one-dimensional bending energy $S_2$, which are defined by
\begin{equation}
\label{Disc-Eneg} 
S_1=\sum_{(ij)} \left(X_i-X_j\right)^2,\quad S_2=\sum^\prime_{(ij)}\left(1-{\bf t}_i\cdot {\bf t}_j\right).
\end{equation} 
$ \sum_{(ij)} $ in $S_1$ denotes the sum over bonds $(ij)$, which connect the vertices $i$ and $j$. In $S_2$, ${\bf t}_i$ is a unit tangential vector of the bond $i$. The symbol $\sum^\prime_{(ij)}$ in $S_2$ is defined below.

\begin{figure}[hbt]
\centering
\includegraphics[width=12.0cm]{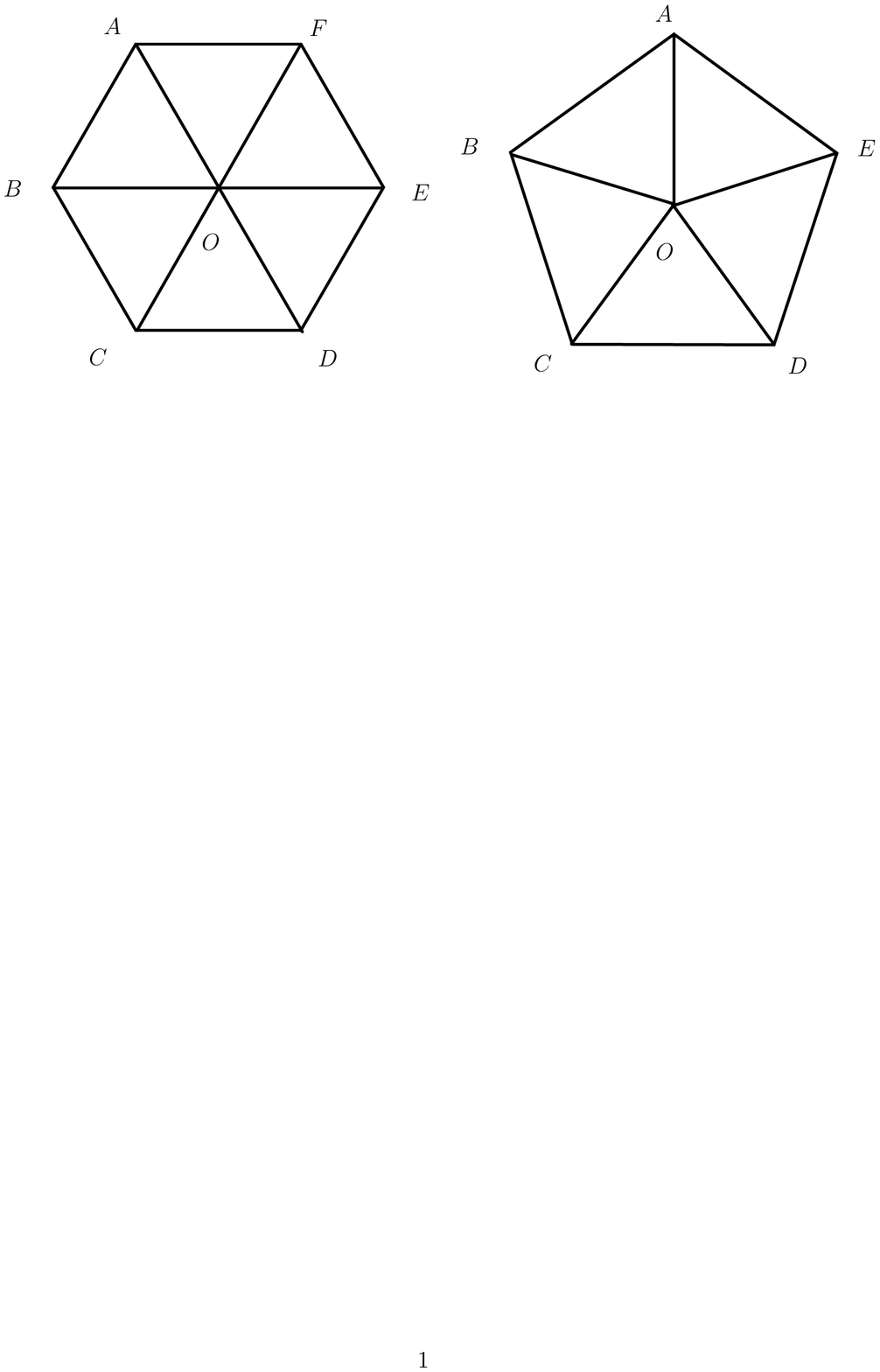}
\caption{(a) A vertex of coordination number $q\!=\!6$, and (b) a vertex of coordination number $q\!=\!5$. The bending energy in (a) is defined by three pairings $1\!-\!{\bf t}_{AO}\cdot {\bf t}_{OD}$, $1\!-\!{\bf t}_{BO}\cdot {\bf t}_{OE}$, and $1\!-\!{\bf t}_{CO}\cdot {\bf t}_{OF}$ at the vertex $O$, and it is defined by the 2.5 pairings at the vertex $O$ in (b) as stated in the text.  }
\label{fig-2}
\end{figure}

The pairing $(ij)$ of the vectors ${\bf t}_i$ and  ${\bf t}_j$ in $S_2$ is defined as follows: At the vertex of coordination number $q\!=\!6$ such as $O$ in Fig.\ref{fig-2}(a), we have three pairings $1\!-\!{\bf t}_{AO}\cdot {\bf t}_{OD}$, $1\!-\!{\bf t}_{BO}\cdot {\bf t}_{OE}$, and $1\!-\!{\bf t}_{CO}\cdot {\bf t}_{OF}$.  The bending energy on the vertex of coordination number $q\!=\!5$ such as $O$ in Fig. \ref{fig-2}(b) is defined by five parings $(1\!-\!{\bf t}_{AO}\cdot {\bf t}_{OC})/2$, $(1\!-\!{\bf t}_{AO}\cdot {\bf t}_{OD})/2$, $(1\!-\!{\bf t}_{BO}\cdot {\bf t}_{OD})/2$, $(1\!-\!{\bf t}_{BO}\cdot {\bf t}_{OE})/2$, and $(1\!-\!{\bf t}_{CO}\cdot {\bf t}_{OE})/2$. Then, we effectively have $2.5$ parings at the $q\!=\!5$ vertices because of the factor $1/2$; $\sum^\prime_{(ij)}$ in $S_2$ is defined by $\sum^\prime_{(ij)}{\bf 1} $, where ${\bf 1}\!=\!1 $ at the vertices of $q\!=\!6$ and ${\bf 1}\!=\!1/2$ at the vertices of $q\!=\!5$. Thus, we have $\sum^\prime_{(ij)}{\bf 1} \!=\!N_B$, where $N_B\!=\!3N-6$ is the total number of bonds.

The partition function of the model is defined by
\begin{eqnarray} 
\label{Part-Func}
 Z = \int^\prime \prod _{i=1}^{N} d X_i \exp\left[-S(X)\right],\\  
 S(X)=S_1 + b S_2, \nonumber
\end{eqnarray}
where $\int^\prime$ denotes that the center of the surface is fixed in the three-dimensional integrations $\int^\prime\prod _{i=1}^{N} d X_i$, where $X_i$ is the three dimensional position of the vertex $i$. Because of the scale invariance of $Z$, $S_1$ is expected to be $S_1/N\!=\!3/2(N\!-\!1)/N\!\simeq\!3/2$. 

The canonical Monte Carlo (MC) technique is used to simulate the multiple three-dimensional integrations in $Z$. The position $X$ is shifted to $X^\prime\!=\!X\!+\!\delta X$, where $\delta X$ is a position chosen randomly in a small sphere. The acceptance rate of the new position is about $50\%$. The radius of the small sphere is fixed at the beginning of the simulations. A random number sequence called Mersenne Twister \cite{Matsumoto-Nishimura-1998} is used to the three-dimensional random shift and the Metropolis accept/reject in the MC simulations. 

The total number of Monte Carlo sweeps (MCS) at the region of the transition point after the thermalization MCS is $9\times 10^8\sim 8\times 10^8$ for the $N\!=\!21162$ and $N\!=\!15212$ surfaces,  $6\times 10^8\sim 5\times 10^8$ for the $N\!=\!10242$ surface, $4\times 10^8\sim 3\times 10^8$ for the $N\!=\!7292$ surface, and $3\times 10^8\sim 2\times 10^8$ for the $N\!=\!4842$ and $N\!=\!2562$ surfaces. Relatively small number of MCS is performed at non-transition region of $b$ in each $N$.

\section{Results}
\begin{figure}[hbt]
\vspace{0.5cm}
\unitlength 0.1in
\begin{picture}( 0,0)(  20,20)
\put(22,56.5){\makebox(0,0){(a)}}%
\put(43,56.5){\makebox(0,0){(b)}}%
\put(22,18.5){\makebox(0,0){(c)}}%
\put(43,18.5){\makebox(0,0){(d)}}%
\end{picture}%
\vspace{0.5cm}
\centering
\includegraphics[width=10.5cm]{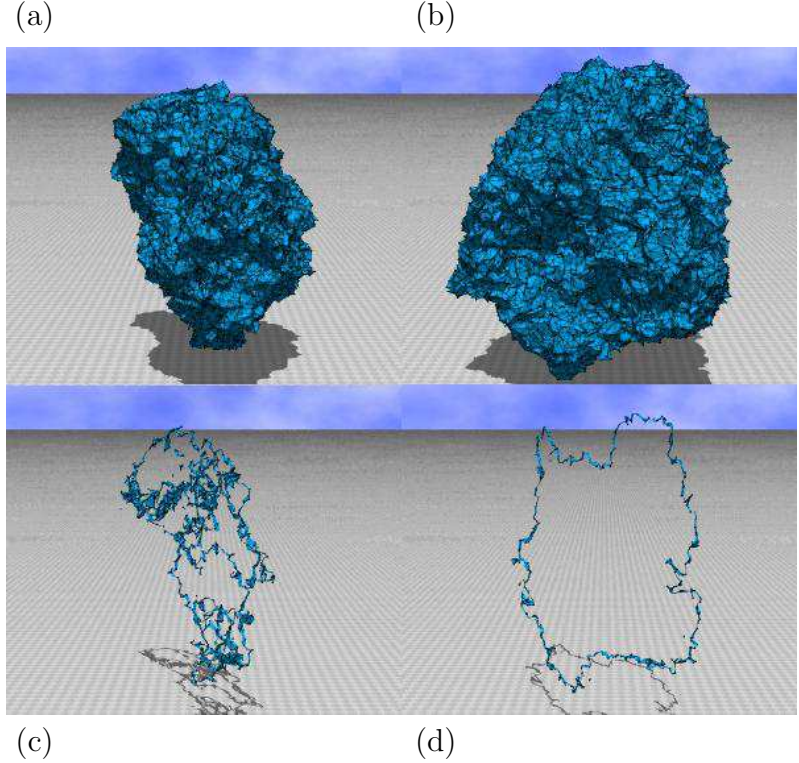}
\caption{Snapshot of surfaces of size $N\!=\!21162$ obtained at $b\!=\!0.718$ in (a) the collapsed phase and (b) the smooth phase. (c) The section of the surface in (a), and (d) the section of the surface in (b). The mean square size $X^2$ is about $X^2\!=\!79$ in (a) and $X^2\!=\!145$ in (b). }
\label{fig-3}
\end{figure}

Snapshots of surface of size $N\!=\!21162$ are shown in Figs.\ref{fig-3}(a) and \ref{fig-3}(b), which were obtained at $b\!=\!0.718$ in the collapsed phase and in the smooth phase, respectively. Figures \ref{fig-3}(c) and \ref{fig-3}(d) show the surface sections in Figs.\ref{fig-3}(a) and \ref{fig-3}(b). These four figures are shown in the same scale. The mean square size $X^2$ is about $X^2\!=\!79$ in (a) and $X^2\!=\!145$ in (b). 

The mean square size $X^2$ is defined by
\begin{equation}
\label{X2}
X^2={1\over N} \sum_i \left(X_i-\bar X\right)^2, \quad \bar X={1\over N} \sum_i X_i,
\end{equation}
where $\bar X$ is the center of mass of the surface. $X^2$ is expected to reflect the size or the shape of surfaces whenever the model has a smooth swollen phase and a collapsed phase. 

\begin{figure}[hbt]
\centering
\includegraphics[width=12.5cm]{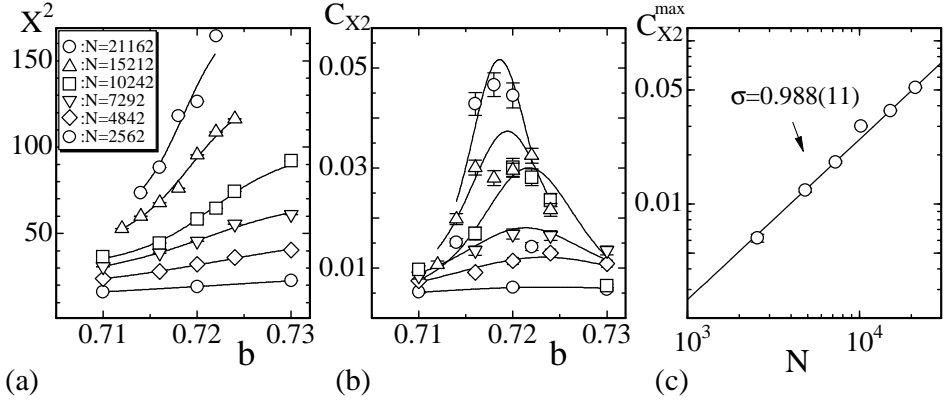}
\caption{(a) The mean square size $X^2$ versus $b$, (b) the variance $C_{X^2}$ of $X^2$ versus $b$, and (c) log-log plots of the peak values $C_{X^2}^{\rm max}$ against $N$. The solid lines in (a), (b) and the data $C_{X^2}^{\rm max}$ in (c) were obtained by the multihistogram reweighting technique. }
\label{fig-4}
\end{figure}
 Figure \ref{fig-4}(a) shows $X^2$ versus $b$. The solid lines were obtained by the multihistogram reweighting technique. The variation of $X^2$ appears smooth against $b$, although it becomes rapid with increasing $N$. The variance $C_{X^2}$ of $X^2$ defined by 
\begin{equation}
\label{fluctuation-X2}
C_{X^2} = {1\over N} \langle \; \left( X^2 \!-\! \langle X^2 \rangle\right)^2\rangle
\end{equation}
is plotted in Fig.\ref{fig-4}(b) against $b$. We clearly see in $C_{X^2}$ an anomalous peak, which grows with increasing $N$. The anomalous peak seen in $C_{X^2}$ represents a collapsing transition between the smooth swollen phase and the collapsed phase.

 In order to see the order of the transition, we plot the peak values $C_{X^2}^{\rm max}$ in Fig.\ref{fig-4}(c) in a log-log scale against $N$. The peak values $C_{X^2}^{\rm max}$ and the statistical errors were obtained also by the multihistogram reweighting technique. The straight line in Fig.\ref{fig-4}(c) was drawn by fitting the data to the scaling relation
\begin{equation}
\label{scaling-CX2}
C_{X^2}^{\rm max} \propto N^\sigma,
\end{equation}
where $\sigma$ is a scaling exponent. The fitting was done by using the data plotted in Fig.\ref{fig-4}(c) excluding that of $N\!=\!10242$.  Thus, we have
\begin{equation}
\label{exponent-CX2}
\sigma = 0.988 \pm 0.011,
\end{equation}
which indicates that the collapsing transition is of first-order. The finite-size scaling (FSS) theory predicts that a transition is of first-order (second-order) if the exponent satisfies $\sigma\!=\!1$ ($\sigma\!<\!1$). 

\begin{figure}[hbt]
\centering
\includegraphics[width=11.5cm]{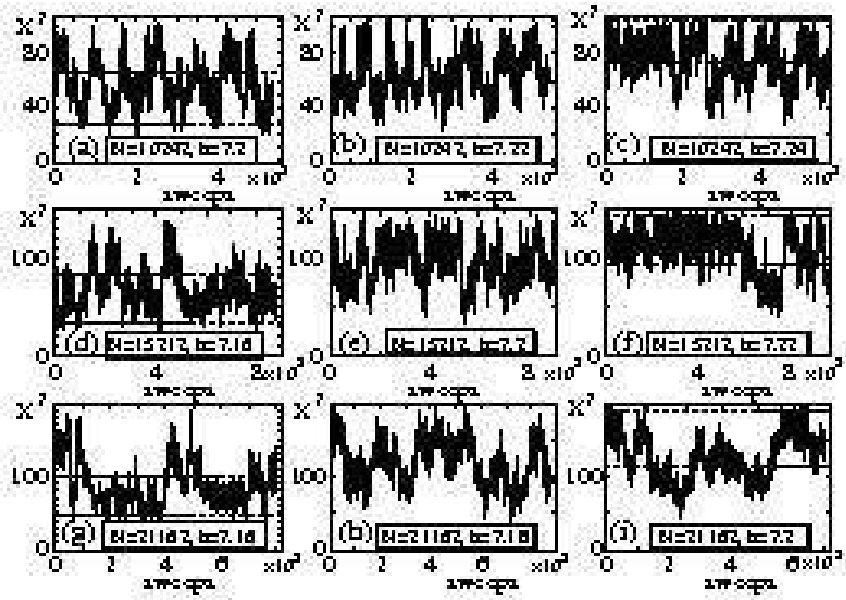}
\caption{The variation of $X^2$ against MCS on the surface of size (a), (b), (c) $N\!=\!10242$,  (d), (e), (f) $N\!=\!15212$, and (g), (h), (i) $N\!=\!21162$. The data were obtained at three distinct $b$ close to the transition point in each $N$, where those in (b), (e), and (h) are considered to be the ones obtained at the transition point.  Horizontal dashed lines in the figures denote $X^2_{\rm min}$ and $X^2_{\rm max}$, which are shown in Table \ref{table-1}.}
\label{fig-5}
\end{figure}

The Hausdorff dimension $H$ is defined by $X^2\sim N^{2/H}$. We expect that $H\simeq 2$ is satisfied in the smooth phase, whereas the value of $H$ in the collapsed phase is unclear, because $X^2$ smoothly changes at the transition point as we see in Fig.\ref{fig-4}(a). Therefore, in order to see the behavior of $X^2$ at the transition point more clearly, we plot the variation of $X^2$ against MCS in Figs.\ref{fig-5}(a)--\ref{fig-5}(i). The variations were obtained at $b\!=\!7.2$, $b\!=\!7.22$, $b\!=\!7.24$ on the $N\!=\!10242$ surface, at $b\!=\!7.16$, $b\!=\!7.2$, $b\!=\!7.22$ on the $N\!=\!15212$ surface, and at $b\!=\!7.16$, $b\!=\!7.18$, $b\!=\!7.2$ on the $N\!=\!21162$ surface.

 We find from the figures that the value of $X^2$ at the smooth phase is not so clearly separated from that of the collapsed phase at the transition point. In fact, we can see a double peak structure only in the histogram $h(X^2)$ of $X^2$ in Fig.\ref{fig-5}(h), although the double peaks are not so clear in the histogram, which is not depicted as a figure. No double peak structure was seen in $h(X^2)$ on the surfaces of $N\!\leq \! 15212$. 

However, the mean value of $X^2$ at the smooth phase and that at the collapsed phase can be obtained from the series of $X^2$ in Figs.\ref{fig-5}(a)--\ref{fig-5}(i) by averaging $X^2$ between the lower bound $X^2_{\rm min}$ and the upper bound  $X^2_{\rm max}$ assumed in each phase. Horizontal dashed lines in the figures denote $X^2_{\rm min}$ and $X^2_{\rm max}$.

The assumed values of $X^{2 \;{\rm col}}_{\rm min}$ and $X^{2 \;{\rm col}}_{\rm max}$in the collapsed phase and those of $X^{2 \;{\rm smo}}_{\rm min}$ and $X^{2 \;{\rm smo}}_{\rm max}$ in the smooth phase are shown in Table \ref{table-1}. The symbols ${\rm col}$ and  ${\rm smo}$ denote the collapsed phase and the smooth phase, respectively. The values $X^2$ in the collapsed phase on the surfaces of $N\!=\!21162$, $N\!=\!15212$, $N\!=\!10242$, $N\!=\!7292$ were respectively obtained at $b\!=\!7.16$, $b\!=\!7.16$, $b\!=\!7.2$,  $b\!=\!7.16$. On the other hand, those $X^2$ in the smooth phase on the surfaces of $N\!=\!21162$, $N\!=\!15212$, $N\!=\!10242$, $N\!=\!7292$ were respectively obtained at $b\!=\!7.2$, $b\!=\!7.22$, $b\!=\!7.24$, $b\!=\!7.24$.
\begin{table}[hbt]
\caption{ The assumed values of the lower bound $X^{2 \;{\rm col}}_{\rm min}$ and the upper bound $X^{2 \;{\rm col}}_{\rm max}$ for obtaining the mean value $X^2$ in the collapsed phase close to the transition point, and those $X^{2 \;{\rm smo}}_{\rm min}$ and $X^{2 \;{\rm smo}}_{\rm min}$ in the smooth phase close to the transition point. $b$(col) and $b$(smo) denote the bending rigidities where $X^2$ was obtained. }
\label{table-1}
\begin{center}
\begin{tabular}{ccccccc}
\hline
$N$ & $b$(col) & $X^{2 \;{\rm col}}_{\rm min}$ & $X^{2 \;{\rm col}}_{\rm max}$ & $b$(smo) &$X^{2 \;{\rm smo}}_{\rm min}$ & $X^{2 \;{\rm smo}}_{\rm max}$   \\
 \hline
21162  &  7.16  & 45  &   100   & 7.2    &   115  &   190  \\
15212  &  7.16  & 35  &    85   & 7.22   &    95  &   145  \\
10242  &  7.2   & 28  &    65   & 7.24   &    73  &   104  \\
 7292  &  7.16  & 20  &    41   & 7.24   &    55  &    78  \\
 \hline
\end{tabular} 
\end{center}
\end{table}

\begin{figure}[hbt]
\centering
\includegraphics[width=5.0cm]{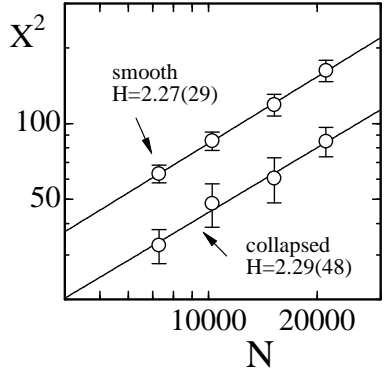}
\caption{Log-log plots of $X^2$ against $N$ obtained in the smooth phase and in the collapsed phase close to the transition point. The straight lines were drawn by fitting the data to $X^2\sim N^{2/H}$, where $H$ is the Hausdorff dimension.
 }
\label{fig-6}
\end{figure}
Figure \ref{fig-6} shows log-log plots of $X^2$ versus $N$ obtained in the smooth phase and in the collapsed phase. The straight lines were obtained by fitting the data to the relation $X^2\sim N^{2/H}$, and we have the Hausdorff dimensions $H_{\rm smo}$ and $H_{\rm col}$ respectively in the smooth phase and in the collapsed phase such that 
\begin{equation}
\label{H}
H_{\rm smo} = 2.27 \pm 0.29, \quad H_{\rm col} = 2.29 \pm 0.48.
\end{equation}
The value of $H_{\rm smo}$ is consistent to the expectation from the snapshot in Figs.\ref{fig-3}(b) and \ref{fig-3}(d). Moreover, we find from $H_{\rm col}$ in Eq.(\ref{H}) that the collapsed phase is considered to be physical, although  $H_{\rm col}$ includes a large error. We must note that these values of $H$ are dependent on the lower and the upper bounds $X^2_{\rm min}$ and $X^2_{\rm max}$, and therefore the results in Eq.(\ref{H}) are not so conclusive. Nevertheless, we feel that the phase transition of the model in this Letter is realistic. The physical condition $H_{\rm col}\!<\!3$ is expected to be obtained more conclusively by large scale simulations.

\begin{figure}[hbt]
\centering
\includegraphics[width=12.5cm]{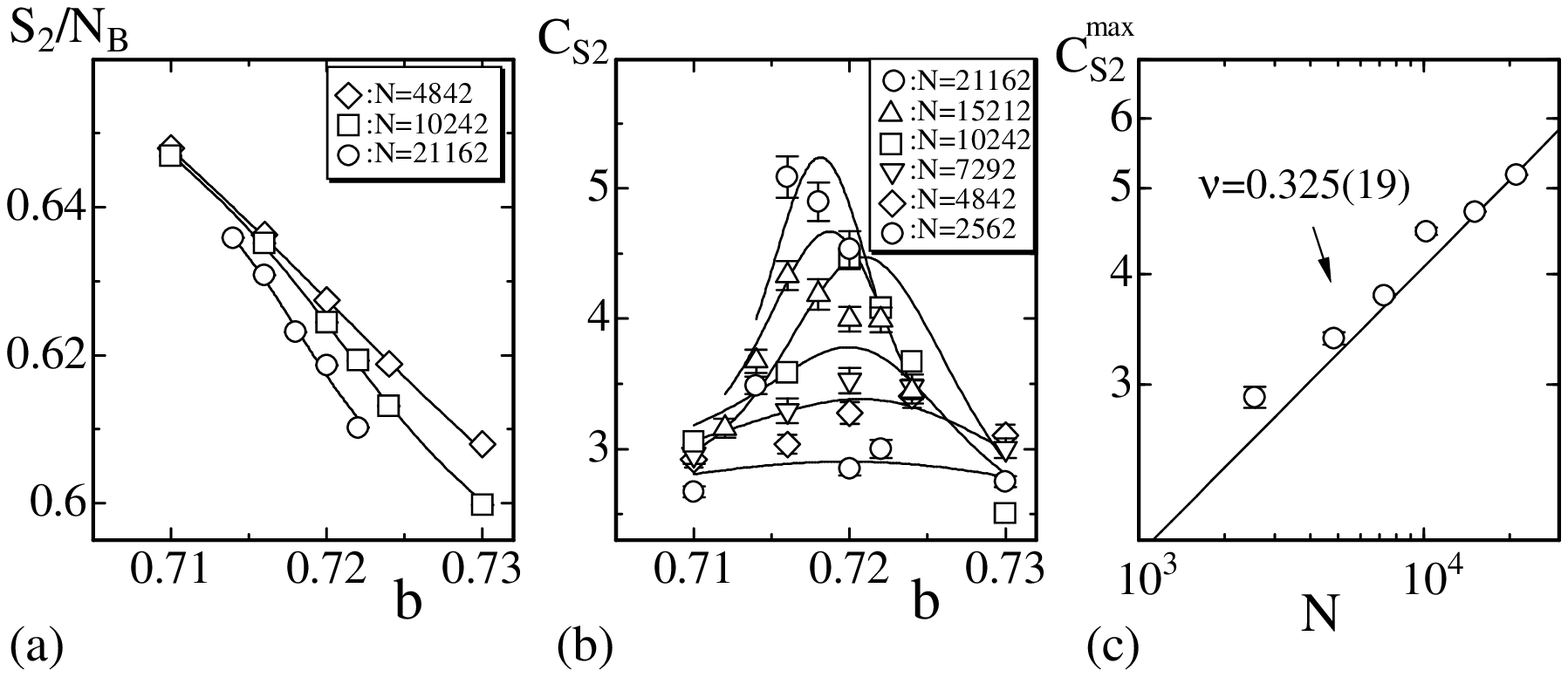}
\caption{(a) The bending energy $S_2/N_B$ versus $b$, (b) the specific heat $C_{S_2}$ versus $b$, and (c) log-log plots of the peak values $C_{S_2}^{\rm max}$ against $N$. The solid lines in (a), (b) and the data $C_{S_2}^{\rm max}$ in (c) were obtained by the multihistogram reweighting technique. The straight line in (c) was drawn by fitting the data to $C_{S_2}^{\rm max} \propto N^\nu$. }
\label{fig-7}
\end{figure}
Figure \ref{fig-7}(a) shows the bending energy $S_2/N_B$ versus $b$ on the surface size $N\!=\!4842$, $N\!=\!10242$, and $N\!=\!21162$. The reason for dividing $S_2$ by $N_B$ is that $\sum^\prime_{(ij)}$ in $S_2$ of Eq.(\ref{Disc-Eneg}) satisfies  $\sum^\prime_{(ij)}{\bf 1}\!=\!N_B$ as mentioned in the previous section. The slope of $S_2/N_B$ becomes large with increasing $N$ as expected. 

The specific heat $C_{S_2}$ defined by
\begin{equation}
C_{S_2} = {b^2\over N} \langle \; \left( S_2 \!-\! \langle S_2 \rangle\right)^2\rangle
\end{equation}
is plotted in Fig.\ref{fig-7}(b). An anomalous peak can also be seen in $C_{S_2}$ at the same transition point as that of the peak of $C_{X^2}$ in Fig.\ref{fig-4}(b). The peak values $C_{S_2}^{\rm max}$ are shown in Fig.\ref{fig-7}(c) in a log-log scale against $N$. We draw in Fig.\ref{fig-7}(c) the straight line which is obtained by the least squares fitting with the inverse statistical errors. The scaling relation is given by $C_{S_2}^{\rm max} \propto N^\nu$, and we have $\nu\!=\!0.325\pm 0.019$. Thus, we understand that the surface fluctuation corresponding to the fluctuation of $S_2$ is a phase transition and is of second-order because of the argument of the FSS theory.

\begin{figure}[hbt]
\centering
\includegraphics[width=12.5cm]{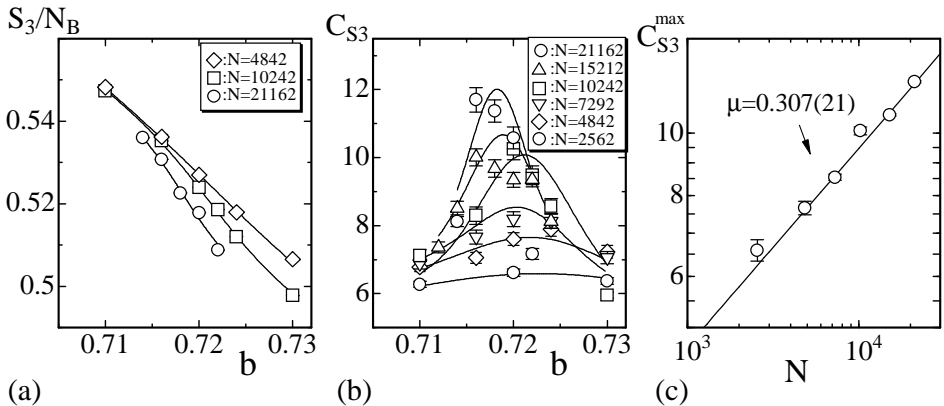}
\caption{(a) The bending energy $S_3/N_B$ versus $b$, (b) the specific heat $C_{S_3}$ versus $b$, and (c) log-log plots of the peak values $C_{S_3}^{\rm max}$ against $N$. The solid lines in (a), (b) and the data $C_{S_3}^{\rm max}$ in (c) were obtained by the multihistogram reweighting technique. The straight line in (c) was drawn by fitting the data to $C_{S_3}^{\rm max} \propto N^\mu$.}
\label{fig-8}
\end{figure}
The standard two-dimensional bending energy is defined by $S_3\!=\!\sum(1\!-\!{\bf n}_i\cdot {\bf n}_j)$, where ${\bf n}_i$ is the unit normal vector of the triangle $i$. The bending energy $S_3$ is expected to reflect the surface fluctuations, although it is not included in the Hamiltonian.
Figure \ref{fig-8}(a) shows $S_3/N_B$ versus $b$, where the surface size is $N\!=\!4842$, $N\!=\!10242$, and $N\!=\!21162$. The variance $C_{S_3} \!=\! {1\over N} \langle \; \left( S_3 \!-\! \langle S_3 \rangle\right)^2\rangle$ defined by the expression similar to that of $C_{X^2}$ in Eq.(\ref{fluctuation-X2}) is plotted in Fig.\ref{fig-8}(b), and the peaks $C_{S_3}^{\rm max}$ obtained  by the the multihistogram reweighting technique are plotted against $N$ in  Fig.\ref{fig-8}(c) in a log-log scale. The straight line in Fig.\ref{fig-8}(c) was obtained by the least squares fitting, which was performed by using all the data in Fig.\ref{fig-8}(c). Thus, we have a scaling exponent $\mu$ in the relation $C_{S_3}^{\rm max}\!\propto \! N^\mu$ such that $\mu\!=\!0.307\pm 0.021$. This result indicates that the surface fluctuation transition is of second-order.

\begin{figure}[hbt]
\centering
\includegraphics[width=12.5cm]{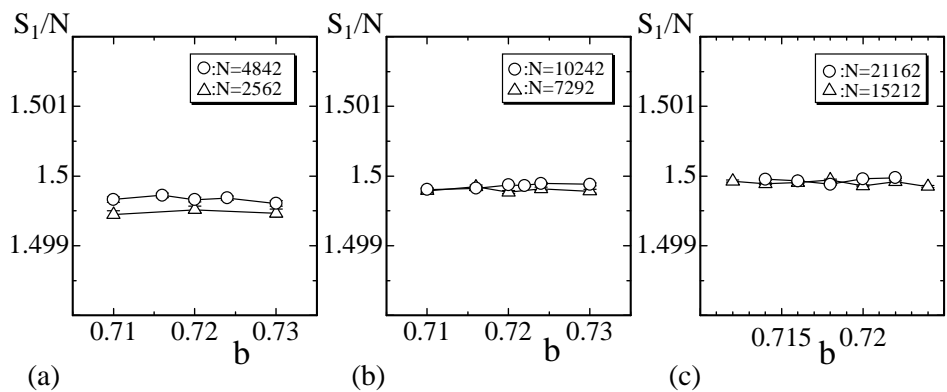}
\caption{The Gaussian bond potential $S_1/N$ versus $b$ obtained on the toroidal surfaces of (a) $N\!=\!2562$, $N\!=\!4842$, (b)  $N\!=\!7292$, $N\!=\!10242$, and (c) $N\!=\!15212$, $N\!=\!21162$. The expected relation $S_1/N\!\simeq\!1.5$ is satisfied.}
\label{fig-9}
\end{figure}
Finally, we plot in Figs.\ref{fig-9}(a)--\ref{fig-9}(c) the Gaussian bond potential $S_1/N$ against $b$. As mentioned in the previous section, $S_1/N$ is expected to be $S_1/N\!\simeq\!1.5$ because of the scale invariant property of the partition function and that of $S_2$. This relation can always be used to check that the simulations were performed successfully. We see in the figures that the expected relation is satisfied. 

\section{Summary and conclusions}
A triangulated surface model has been investigated by using the Monte Carlo simulation technique. Hamiltonian of the model is given by a linear combination of the Gaussian bond potential and a one-dimensional bending energy. The model is considered to be obtained from a compartmentalized surface model in the limit of $n\!\to\!0$, where $n$ is the total number of vertices in a compartment and hence denotes the size of compartment. 
 
We have found that the model in this Letter undergoes a first-order collapsing transition and a second-order surface fluctuation transition. On the other hand, we know that the compartmentalized model with the two-dimensional elasticity at the junctions undergoes a first-order surface fluctuation transition \cite{KOIB-JSTP2007}, moreover a compartmentalized fluid surface model with the rigid junction also undergoes a first-order one \cite{KOIB-PRE2007-2}. Therefore, we consider that the fluctuation of vertices inside the compartments strengthen the surface fluctuation transition in the $n\!\not=\!0$ model. On the contrary, we have no vertices inside the compartments in the model of this Letter because of $n\!=\!0$. The lack of vertex fluctuation is considered to soften the first-order surface fluctuation transition seen in the finite $n$ model.  

We should note that sufficiently small values of $n$ implies that the compartment size is comparable to the bond length scale, which can arbitrarily be fixed due to the scale invariant property of the partition function. The size $n$ is proportional to the area of a compartment, and hence the finite $n$ implies that the corresponding compartment size is negligible compared to the surface size in the limit of $N \to \infty$. The finite $n$ also implies that the compartment size is sufficiently larger than the bond length scale. Thus, the model in this Letter is considered to be a compartmentalized model with sufficiently small compartment. 

The model in this Letter is allowed to self-intersect and hence phantom. A phantom surface model, which has a collapsing transition between the smooth phase and the collapsed phase, is considered to be realistic if the collapsed phase is physical. One of the criteria for such physical condition is given by $H \!<\! 3$, where $H$ is the Hausdorff dimension. Therefore, in order to see whether the condition is satisfied or not in our model, we obtained $X^2$ in the smooth phase and in the collapsed phase close to the transition point by averaging $X^2$ between $X^2_{\rm min}$ and $X^2_{\rm max}$ assumed in each phase. Thus, $H_{\rm smo} \!=\! 2.27(29)$ (smooth phase) and $H_{\rm col} \!=\! 2.29(48)$ (collapsed phase) were obtained, and then we found that the physical condition  $H \!<\! 3$ is satisfied in the collapsed phase although $H_{\rm col} $ includes relatively large error.

Meshwork models in \cite{KOIB-LNCS2006,KOIB-JST-2007-2} has no vertex inside the compartments, which have finite size $n$. The phase structure of such meshwork model of finite $n$ is considered to be dependent on the elasticity of junctions \cite{KOIB-LNCS2006,KOIB-JST-2007-2}. Therefore, it is interesting to study the dependence of the surface fluctuation transition on $n$ in the meshwork model, where the elasticity of junctions is identical to that in the model of this Letter.

\section*{Acknowledgment}
This work is supported in part by a Grant-in-Aid for Scientific Research from Japan Society for the Promotion of Science.  




\end{document}